\documentstyle[prl,aps,preprint]{revtex}
\def\fl{f\kern-0.02eml\kern-0.05em}
\begin{document}
\title
{The Mean Field Theory for\\
Percolation Models of the Ising Type}
\author{L.\ Chayes}
\address{Department of Mathematics,\\ University of California,
Los Angeles, CA 90095-1555, USA}
\author{A.\ Coniglio}
\address{Dipartimento di Scienze Fisiche,\\ Universit\`a``Federico
II''Mostra d'Oltremare
Pad.19 I--80125 Napoli, Italy}
\author{J.\ Machta}
\address{Department of Physics and Astronomy,\\ University of
  Massachusetts, Amherst, MA 01003-3720, USA}
\author{K.\ Shtengel}
\address{Department of Physics,\\ University of California,
Los Angeles, CA 90095-1547, USA}
\maketitle
\begin{abstract}
\hsize = 6.75 in
\baselineskip = 12pt  

The $q=2$ random cluster model is studied in the context of two mean--field
models:
The Bethe lattice and the complete graph.  For these systems, the critical
exponents that
are defined in terms of finite clusters have some anomalous values as the
critical point
is approached from the high--density side which vindicates the results of
earlier
studies.  In particular, the exponent $\tilde \gamma^\prime$ which
characterises the
divergence of the average size of finite clusters is 1/2 and
$\tilde\nu^\prime$, the exponent associated with the length scale of finite
clusters is
1/4.  The full collection of exponents indicates an upper critical
dimension of 6.
The standard mean--field exponents of the Ising system are also present in
this model ($\nu^\prime = 1/2$, $\gamma^\prime = 1$) which implies, in
particular, the
presence of two diverging length--scales.  Furthermore, the
finite cluster exponents are stable to the addition of disorder which, near
the upper
critical dimension, may have interesting implications concerning the
generality of the
disordered system/correlation length bounds.

\end{abstract}
\newpage
\hsize = 7 in
\subsection {Introduction}
The close connection between spin--systems and
percolation models has lead to many important developments in
statistical physics.  For a broad class of spin models there is a
mapping to an equivalent {\em graphical representation} having a
percolation transition corresponding to the phase transition of the
spin model (see e.g.~\cite{CM}).  In particular, the graphical
representations of the $q$-state Potts models are the ``$q$-state'' random
cluster models~\cite{FK,CK}.  For $q=2$ the
spin representation is the Ising model while the
$q = 1$ random cluster model is ordinary bond percolation.  The
equivalence between spin models and graphical models has lead
to a new class of highly efficient {\em cluster} Monte Carlo
methods~\cite{SW}.  Cluster methods simulate both the spin model and
the graphical representation and it is often more efficient to
measure thermodynamic quantities via their graphical
analogs~\cite{Sweeny}.

As an example of the correspondence between spin models and graphical
models, consider the order parameter exponent $\beta$~\cite{proviso} that is
defined for spin systems by, $m \sim (T_c-T)^\beta$ where $T$ is the
temperature, $T_c$ the critical temperature and $m$ the order
parameter. For graphical models, the order parameter
is the fraction of sites in the percolating cluster, $P_\infty$ and
$\tilde{\beta}$ is defined as, $P_{\infty} \sim
(p-p_c)^{\tilde{\beta}}$ where $p$ is the bond occupation
probability and $p_c$ the percolation threshold.  For a given
$q$-state Potts model and the corresponding
random cluster model it turns out that
$m=P_{\infty}$ so that $\tilde{\beta}=\beta$.  Analogs of  other
thermal exponents may be defined for percolation models;
$\tilde{\gamma}$ ($\tilde{\gamma}^\prime$) characterise the
divergence of the average size of the {\em finite} cluster
containing the origin below (above) the percolation threshold.
Similarly, $\tilde{\nu}$ ($\tilde{\nu}^\prime$) characterises the
divergence of average scale of the {\em finite}
clusters below (above) threshold.  The connection between the thermal
and geometric exponents follows from the fact that the spin-spin
correlation function between sites $i$ and $j$ is equal to the
probability that $i$ and $j$ are in the same cluster in this graphical
representation.  Thus, in the high temperature phase the correlation lengths for
the two models are equal and the magnetic susceptibility is equal to the average
size of the connected clusters.  Thus, in particular,
$\gamma = \tilde{\gamma}$ and $\nu = \tilde{\nu}$.

On the other hand, the corresponding relation for
${\gamma}^\prime$  and $\tilde{\gamma}^\prime$
or (${\nu}^\prime$ and $\tilde{\nu}^\prime$) is {\em not} a straightforward
consequence of the mapping to the graphical representation.  In particular, for
the Potts models, the graphical expression for the truncated two point function
between the sites $i$ and $j$ consists of two terms: (i) The probability that
the two sites are in the same finite cluster and (ii) the correlation of the
infinite cluster density at the sites $i$ and $j$.  For the random cluster
models (with $q \geq 1$) the second term is positive which leads to the
inequalities $\tilde{\nu} \geq \tilde{\nu}^\prime$ and $\tilde{\gamma} \geq
\tilde{\gamma}^\prime$.  On general grounds, one would expect that items (i) and
(ii) are comparable and for independent percolation ($q=1$) this is the
subject of a rigorous theorem~\cite{CCGKS}.  It is therefore something of a
surprise that for percolation models of the Ising--type, the equality of
$\tilde{\gamma}$ and $\tilde{\gamma}^\prime$ breaks down in mean--field.
This was first discovered in a series of papers co--authored by one of us
\cite{CoStKl,RoCoSt,CoLiMoPe} both in the context of a polymer/solvent model
and in a mean--field calculation for the Potts/percolation droplet model.

The principal focus of this paper is to underscore (and bolster) the
conclusions inherent in the result ${\gamma}^\prime \neq \tilde{\gamma}^\prime$
and to clarify the relationship between the Ising magnet and its graphical
representation.  In particular, the result
${\gamma}^\prime \neq \tilde{\gamma}^\prime$ in these sorts of systems will be
shown in some generality:  A full $\text{\fl}$edged Bethe lattice
calculation and for the
$q=2$ random cluster model defined on the complete graph.  Furthermore (at least
in the context of the Bethe lattice) there is the additional mean--field
exponent
$\tilde{\nu}^\prime = \frac 14$.  Taken along with $\tilde\beta =\beta =
\frac 12$ and $\eta = 0$,
these exponents satisfy the standard scaling relations and lead to the
tentative prediction $d_c = 6$ for the upper critical dimension.
Of particular interest is therefore the behaviour
of the $q = 2$ random cluster model itself:  a different set of exponents
above and
below threshold and the appearance of two distinct diverging length scales as
the critical point is approached from the percolating phase. Since these
models are
the starting point of the $\infty > d \gg 1$ expansions, it is our belief
that the behaviour uncovered
here also holds in sufficiently high dimensions.

It must be emphasised that the
discrepancies between random cluster and Ising exponents are due
to a difference in definitions not content.  The same information is
inherent in both a spin model and its graphical representation and
it is always possible to define appropriate graphical quantities
to represent any spin quantity.  On the other hand, the
fact that two natural sets of definitions lead to different
exponents indicates the presence of real and interesting features of
the Ising--type random cluster model in high dimensions.  Some of these
are discussed at the conclusion of this paper.

\subsection {The Bethe Lattice}
Consider a half-space Bethe lattice of coordination number 3
(a binary Cayley tree). For a $k$-level tree define the interacting
bond percolation problem with the weights of configurations $\omega$
given by $W(\omega )=B_{p_e}(\omega )\, q^{l(\omega )}$. Here $
B_{p_e}(\omega ) $ is the Bernoulli probability of $ \omega $ at the
parameter value $ p_e \in [0,1]$. The quantity $l(\omega )$ counts the
number of loops. In the present context, the only mechanism for the
formation of loops is via connections to the boundary; all points
on the boundary are regarded as ``pre-connected" \cite{Chayes 1}.
As
is well known (for appropriate $q$) this is the random cluster
representation the $q$-state Potts model given by the
Hamiltonian $ - \beta H = J\sum_{<i,j>}{\delta_{\sigma_i,\sigma_j}}$
subject to the boundary condition that the boundary sites are all
locked in the same state.  The parameter $p_e$ is given by $ p_e=
[1-e^{-J}]/[1+(q-1)e^{-J}] $ (which for the Ising case $q=2$ becomes
$ p_e=\mbox{tanh}{(J/2)} $).

  The partition function on the $k$-level binary tree is
given by $Z_{k}\equiv \sum _{\omega }W_k(\omega)$.  There are two
types of bond configurations: those in
which the root site is connected to the boundary, and those in which
the root site is not. Let $ I_{k} $ denote the total weight of
configurations connected to the boundary and $F_{k}$ denote the
total weight of bond configurations not connected to the boundary; $
Z_{k}=I_{k}+F_{k} $.  Let us now take two such $k$-level trees and
join them by attaching their root sites to a new root site, thus
creating a $ (k+1) $-level tree.  The recursion relations for $
F_{k+1} $, $ I_{k+1}$, and $Z_{k+1}$ are:
\begin{mathletters}
\label{FIZ}
\begin{equation}
F_{k+1} = \left[F_{k}+(1-p_e)I_{k}\right]^2 = \left(Z_{k}-p_e
    I_{k}\right)^2\,;
 \label{F_k}
\end{equation}
\begin{equation}
I_{k+1}  = 2 p_e I_k Z_k + (q-2)p_e^2 I_k^2\,;
 \label{I_k}
\end{equation}
\begin{equation}
Z_{k+1} = Z_k^2+(q-1)p_e^2 I_k^2\,.
 \label{Z_k}
\end{equation}
\end{mathletters}

We define $P_{\infty}^{(k)} \equiv {I_{k}}/{Z_{k}}$ to be the
probability that a root site of a $k$-level tree is connected to the
boundary.  Using the above recursion relations (\ref{FIZ}) and taking
the limit of $k \rightarrow \infty$, we arrive at the well-known
expression for the percolation density:
\begin{equation}
  P_{\infty}\equiv \lim_{k \rightarrow \infty}P_{\infty}^{(k)} =\frac
  {2 p_e P_{\infty} + (q-2)p_e^2 P_{\infty}^2 }{1+(q-1)p_e^2
    P_{\infty}^2 }\,.
\label{Pinf}
\end{equation}
Analysing the non-trivial solution of this equation one can see that
there is a continuous phase transition at $p_e=1/2$ if $0< q \leq
2$ with the critical exponent $\tilde{\beta} = 1$ for $0< q < 2$ and
$\tilde{\beta} = 1/2$ for $q=2$. For $q>2$ the transition becomes
discontinuous.

We now look at the average {\em finite} cluster size.  Let us define a
random variable $c_k$ to be the size of the connected component of the
root site, whenever this cluster is disconnected from the boundary,
and $c_k = 0$ if the root site is connected to the boundary.  Let also
$X _k \equiv \sum_{\omega}c_k W_k(\omega)/Z_k$ be its average. Since
only the bond configurations that contribute to $F_k$ enter into this
average, we can write $X _k = f_k {F_k}/{Z_k}$ where $f_k$ is the
average cluster size {\em given} that it is disconnected from the
boundary.  Again, merging two $k$-level trees into a new $(k+1)$-level
tree one finds the recursion relation for $f_{k}$:
\begin{eqnarray}
  f_{k+1} {F_{k+1}} = (1-p_e)^2 Z_k^2 + (1+f_k) p_e F_k (1-p_e) Z_k
+ (1+2f_k)p_e^2 F_k^2\,.
\label{clustersize2}
\end{eqnarray}
Dividing both sides of (\ref{clustersize2}) by $Z_{k+1}$
and recalling the definitions of $X _k$ and $P_{\infty}^{(k)}$ we obtain
\begin {equation}
X_{k+1} = 1-P_{\infty}^{(k+1)} +
X_k\frac{2p_e^2(1-P_{\infty}^{(k)}) + 2p_e(1-p_e)}{1 +
(q-1)[p_eP_{\infty}^{(k)}]^2}.
\label{insert}
\end{equation}
Taking
the limit $k \rightarrow \infty$ (the existence of which is guaranteed by a
straightforward argument) we obtain:
\begin{equation}
  X(p_e)\equiv \lim_{k \rightarrow \infty} X _{k} = \frac{(1-p_e
    P_{\infty})^2}{1-2p_e +2p_e^2 P_{\infty} +(q-1)p_e^2
    P_{\infty}^2}\,.
\label{clustersize4}
\end{equation}

This expression is valid on both the ordered and disordered
sides of the transition.  Indeed, for $p_e<1/2$,  $P_{\infty}=0$ and
Eq.(\ref{clustersize4}) reduces to the well known expression
$X={1}/({1-2p_e})$ (which gives $\tilde{\gamma}=1$ for all $0< q \leq
2$).

The situation on the ordered side ($p_e>1/2$) is changed by the
presence of the term proportional to $P_{\infty}$ in the denominator.
For $0<q<2$, $P_{\infty}(p_e)$ is asymptotically linear in $(p_e-1/2)$
resulting in the anticipated $\tilde{\gamma}^\prime=1$. However, for
$q=2$, \begin{equation}
P_{\infty}(p_e)\equiv m(p_e)\sim (p_e - \frac 12)^{1/2}\gg (2 p_e-1)
  \label{Magn-Ising}
\end{equation}
for $(p_e-1/2)$ small, so that the terms involving $P_\infty$ are
dominant. Indeed, substituting this expression for $m$ directly into
Eq.~(\ref{clustersize4}) we arrive at
\begin{equation}
X=\frac{1-m}{p_e m}\qquad
\mbox{ for $q=2$ and $p_e>1/2$.}
\label{clustersize5}
\end{equation}
The result for  $q=2$ is
$\tilde{\gamma}^\prime=1/2\neq\gamma$.

Now we look at the finite cluster distribution. The probability that
the root site of a $k$-level tree belongs to a cluster of size $n$
that does not touch the boundary is defined as
\begin{equation}
  P_n^{(k)}=\frac{F_k(n)}{Z_k}=
 \frac{F_k(n)}{F_k+I_k}\,,
\label{Pn1}
\end{equation}
where ${F_k(n)}$ is simply the total weight corresponding to such
event (clearly, $F_k=\sum_{n}{F_k(n)}$). Once again, merging two
$k$-level trees into a new one we find the following relations between
these weights:
\begin{mathletters}
 \label{Fns}
\begin{eqnarray}
  F_{k+1}(1) & = & (1-p_e)^2 Z_{k}^2\,;
  \label{Fn1}\\
 F_{k+1}(n) & = & 2 p_e(1-p_e)Z_{k} F_{k}(n-1)
 +  p_e^2 \sum_{i=1}^{n-2}F_{k}(i)
 F_{k}(n-i-1)\,,\quad(n\neq 1)\,.
  \label{Fn2}
\end{eqnarray}
\end{mathletters}

The next step is to divide both parts of these equations by $Z_{k+1}$
and to arrive (with the help of (\ref{Z_k})) at the expression for
$P_n^{(k+1)}$ in terms of $P_i^{(k)}$ and $P_{\infty}^{(k)}$.  Letting $k
\rightarrow
\infty$
(which is again easily justified) we obtain:
\begin{mathletters}
 \label{Pns}
\begin{eqnarray}
 P_1 & = &  \frac{(1-p_e)^2}{1+(q-1)p_e^2 P_{\infty}^2}\,;
  \label{Pn2}\\
 P_{n} & = & \frac{1}{1+(q-1)p_e^2 P_{\infty}^2}\,
 \Big \{ 2 p_e(1-p_e) P_{n-1}
+  p_e^2 \sum_{i=1}^{n-2}P_i
 P_{n-i-1}\Big \}\,,\qquad(n\neq 1)\,.
  \label{Pn3}
\end{eqnarray}
\end{mathletters}
It should be noted that (\ref{Pn2}) and (\ref{Pn3}) are independent of
$q$ when $P_{\infty}=0$. Thus, below {\em and at} the percolation
threshold ($p_e\leq 1/2$) the finite cluster distribution is
identical to that for the case of independent percolation!  Since
the critical exponents $\tau$ (and $\eta$) are defined at the critical point
they must take on their mean field percolation values
$\tau=5/2$ (and $\eta=0$) for any $q\in (0,2]$.  Since $\delta=3$
for the mean field Ising model, the exponent relation
$\tau=1/\delta-2$ is violated.

The last critical exponent of our interest here is the correlation
length exponent $\tilde{\nu}$. In order to find it we adopt the
standard definition of the correlation length on the Bethe lattice
(see, e.g.,~Ref.~\cite{Grimmett}):
\begin{equation}
  \xi(p_e)=\sqrt{\frac{1}{X(p_e)}\sum_{x}|x|^2\,
    \tau_{0,x}^f(p_e)}
  \label{CL1}
\end{equation}
where $\tau_{0,x}^f(p_e)$ is the probability of the origin being
connected to the site $x$ while {\em not} being connected to the
boundary. The probability $\tau_{0,x}^f$ depends
only on the level $n$ of the Cayley tree that the site $x$ belongs to.
Thus, using the metrics relation $|x|^2=n$, we can rewrite the sum in
(\ref{CL1}) as
\begin{equation}
\sum_{x}|x|^2\,\tau_{0,x}^f=\sum_{n=0}^{\infty}n 2^n\tau_{n}^f=
2\sum_{n=0}^{\infty}(n+1) 2^n\tau_{n+1}^f\;.
  \label{sum}
\end{equation}
Performing the same procedure of merging two $k$-level trees together
to form a $k+1$-level tree and taking a limit of $k\rightarrow \infty$ we
generate a recursion relation for the probability $\tau_{n}^f$:
\begin{equation}
  \tau^f_{n+1}=\tau^f_{n}\frac{p_e(1-p_e P_{\infty})}{1+(q-1)p_e^2
    P_{\infty}^2}\;.
  \label{tau}
\end{equation}
Combining this result with (\ref{sum}) we obtain the desired
expression for the correlation length:
\begin{equation}
  \xi(p_e)=\sqrt{\frac{2 p_e(1-p_e P_{\infty})}{1-2p_e +2p_e^2
      P_{\infty} +(q-1)p_e^2 P_{\infty}^2}}
=\sqrt{\frac{2 p_e X (p_e)}{1-p_e P_{\infty}}}\
 \label{Whatdis}
\end{equation}
with $X (p_e)$ given by Eq.\ (\ref{clustersize4}).  This brings
us to the following relation between the critical exponents:
$\tilde{\nu}=\tilde{\gamma}/2$ on both sides of the transition (in
agreement with the scaling relation $\gamma=2(\nu-\eta)$ since
$\eta=0$), which in turn leads to a surprising result: in
the Ising case ($q=2$) $\nu^\prime  = 1/2$ while
$\tilde{\nu}^\prime = 1/4\,$!

So far we have dealt with the half-space
Bethe lattice where the root site has only two nearest neighbours as
opposed to three for any inner site. We claim, however, that all
full-space quantities experience the same critical behaviour as the
corresponding half-space quantities above.  In fact, they can be
explicitly calculated if we attach the root sites of two identical
trees together thus forming a complete full-space Bethe lattice
(cf.~\cite{Chayes 1}): \begin{mathletters}
 \label{full_lattice}
 \begin{equation}
   {\cal P}_{\infty}=P_{\infty} \frac{1+p_e + (q-2)p_e P_{\infty}}{1 +
     (q-1)p_e P_{\infty}^2}\,;
   \label{P_full}
 \end{equation}
\begin{equation}
{\cal X}(p_e)= X(p_e) \frac{1+p_e -2 p_e P_{\infty}}{1 +
     (q-1)p_e P_{\infty}^2}\,;
   \label{X_full}
 \end{equation}
\begin{equation}
  {\zeta}(p_e)=\xi(p_e) \sqrt{\frac{3(1-p_e P_{\infty})}{2(1+p_e -2 p_e
      P_{\infty})}}\,.
   \label{xi_full}
 \end{equation}
\end{mathletters}
Here ${\cal P}_{\infty}$, ${\cal X}$, and ${\zeta}$ refer to the
isotropic, full-space quantities, which are just non-singular modifications
of the corresponding half-space quantities given by Eqs.\
(\ref{Pinf}), (\ref{clustersize4}) and (14).

\noindent {\bf The disordered case.\ }
The addition of disorder has almost no
effect on the previous set of results.  This fact leads to some interesting
consequences that will be discussed in the final section.  Here we will
demonstrate this stability to disorder confining attention to the exponent
$\tilde
\gamma ^{\prime}$ for the case $q = 2$.  The setup is as follows:  The
couplings are
be given by $(J_{i,j})$ which are identical and independent non--negative
random variables.
The quantity $P_\infty^{(k)}$ is now a random function of these couplings
that obeys the
distributional equation
\begin {equation}
P_\infty^{(k+1)}  =_{d}
\frac {p_{e,L}P_{\infty,L}^{(k)} + p_{e,R}P_{\infty,R}^{(k)}}
{1 + p_{e,L}P_{\infty,L}^{(k)}p_{e,R}P_{\infty,R}^{(k)}}
\end{equation}
with $P_{\infty,L}^{(k)}$ and $P_{\infty,R}^{(k)}$ identical and independent
representing the percolation probabilities for a $k$--level tree and
with $p_{e,L}$ and $p_{e,R}$, distributed according to
$\tanh (J_{i,j}/2)$, representing the effective strength of the bonds connecting
the root site to the two $k$--level trees situated above the root site.
Let $\overline P^{(k)}_\infty$ denote the (quenched) average of
$P^{(k)}_\infty$ and similarly let $\overline p_e$ denote the average of
$p_{e,L}$ or $p_{e,R}$.  It is assumed that the distribution for the $J_{i,j}$
depends on a parameter (e.g.~width, temperature) that can be changed
continuously.
The phase transition occurs at $\overline p_e = 1/2$ and the exponent
$\beta$ is the same
as in the non--random case.

Our analysis begins with the random analog of Eq. (\ref{insert}).  After a
certain
amount of work, the relevant generalisation is seen to be
\begin {equation}
X_{k+1}  =_d 1 - P^{(k+1)}_\infty +
\frac{p_{e,L}X_{k}(1- p_{e,R}P_{\infty,R}^{(k)}) + p_{e,R}X^{\prime}_{k}(1-
p_{e,L}P_{\infty,L}^{(k)})} {1 +
p_{e,L}P_{\infty,L}^{(k)}p_{e,R}P_{\infty,R}^{(k)}}
\end{equation}
Since the $(J_{i,j})$'s are all non--negative, all the terms in the
denominator are
non--negative and we may write
\begin {equation}
X_{k+1}  \leq_d 1 - P^{(k+1)}_\infty +
 p_{e,L}X_{k}(1- p_{e,R}P_{\infty,R}^{(k)})
+ p_{e,R}X^{\prime}_{k}(1- p_{e,L}P_{\infty,L}^{(k)})
\label{blunk}
\end{equation}
Noting that all the relevant quantities in Eq. (\ref{blunk}) are independent
we perform the disorder average to obtain
\begin {equation}
\overline X_{k+1} \leq 1 - \overline P^{(k+1)}_\infty
+ 2\overline p_e \overline X_{k}(1 - \overline p_e \overline P^{(k)}_\infty)
\end{equation}
In the setup with wired boundary conditions and all the $(J_{i,j})$
non--negative, it is
not difficult to show that these average quantities tend to a definite limit as
$k\to\infty$.  Hence
\begin {equation}
\overline X_\infty \leq \frac{1 - \overline P_\infty}{1-2\overline p_e +
 \overline p_e^{2}\overline P_\infty}.
\label{muf}
\end{equation}
Using $\overline P_\infty \sim (\overline p_e - 1/2)^{1/2}$ -- which is not
hard to prove
-- we obtain the (significant) first half: $\tilde \gamma ^\prime\leq 1/2$.

Opposite bounds are obtained by expanding the denominator
\begin{eqnarray}
  X_{k+1}  \geq_d 1 - P^{(k+1)}_\infty + &[p_{e,L}X_{k}(1-
p_{e,R}P_{\infty,R}^{(k)})
+ p_{e,R}X^{\prime}_{k}(1- p_{e,L}P_{\infty,L}^{(k)})]\times\nonumber \\
 &\times [1 - p_{e,L}p_{e,R}P_{\infty,L}^{(k)}P_{\infty,R}^{(k)}]
  \label{FFF}
\end{eqnarray}
Neglecting positive terms and using $p_{e,L}P_{\infty,L}^{(k)} \leq 1$,
$p_{e,R}P_{\infty,R}^{(k)} \leq 1$ we arrive at
\begin {equation}
\overline X_{k+1} \geq_d 1 - \overline P_\infty^{(k+1)} +
2\overline p_{e}\overline X_{k} -
4\overline p_{e}^{2}\overline P_\infty^{(k)}\overline X_{k}
\end{equation}
which leads to a bound similar to Eq. (\ref{muf}) but in the opposite direction.
Hence we conclude $\tilde \gamma^\prime = 1/2$.

\subsection {The Complete Graph}

Similar results can be established for the random cluster model on the
complete graph.  Here, the calculations are as straightforward as the
Bethe lattice, however, the rigorous justification of these
calculations requires some unpleasant analysis.  We
will again be content with the discussion of the exponent
$\tilde{\gamma}^\prime$.

The underlying lattice consists of $N$ sites with bonds of uniform strength
between all pairs.  The weight for a bond configuration $\omega$ is given by
$W(\omega) = B_{p_e(N)}(\omega)q^{l(\omega)} \propto
B_{p_{N}}(\omega)q^{c(\omega)}$ with
$c$ the number of connected components and $p_N$ defined to be
$1-e^{-J/N}$.  C.f.~\cite{BGJ} for a more detailed description of
the random cluster model in this context.
Let $G$ denote the size of the
largest (giant) cluster.  As is not hard to show, the probability of
belonging to this cluster, $G/N$, converges to $m(J)$ where $m(J)$ satisfies
the mean field equation $m = [1-e^{-Jm}]/[1+(q-1)e^{-Jm}]$~\cite{BGJ}.  We
assume
throughout that $ q = 2$ and $J > 2$ so we are in the low--temperature phase.

If $i$ is a site in the graph, let $C_i(N,p_N)$ denote the analog of the
quantity $c_k$ for the Bethe lattice; that is $C_i(N,p_N)$ is zero if
$i$ is connected to the giant cluster and otherwise is the size of the cluster
at $i$.  The strategy will be to fix $G$ and obtain estimates on (the
distribution of) $C_i$.  These estimates are, more or less, the desired result
if the random $G$ is replaced by $mN$.  The large deviation result of
~\cite{BGJ} in essence allows this replacement.

Thus suppose that there are $G$ sites in the giant cluster of an $N$ site graph
with parameter $p$ (not necessarily equal to $p_N$) and consider the cluster of
the origin. For fixed $\epsilon$, let us assume that $|G/N - m| <
\epsilon$ -- otherwise for the upper bounds we will assign $C_0 = N$ and for
the lower bounds, $C_0 = 0$.  We start with the upper bounds.  As before,
let $F_0$
denote the size of the cluster at $0$ given that it is not attached to the
giant cluster.  Given the condition of detachment, the origin and the other $N -
G - 1$ sites act like an autonomous random cluster model subject to the
condition that
no cluster has size exceeding $G$ -- and for the upper bounds, we may
neglect this
condition. We will use the methods introduced in \cite{CoLiMoPe,ES} known
as the
Edwards--Sokal coupling which for complete graphs is particularly easy:
Divide the remaining $N - G$ sites into two groups of $N_1$ and $N_2$
sites.  One of these is identified as plus and the other as minus.  There
can be no bonds between spins of opposite type and bonds between spins of
the same type occur with probability $p_N$.
Taking into account the relevant energetics and combinatorial factor, the
result is seen to be a complete graph Ising problem with
$N-G$ sites and temperature parameter $p = p_N$.
Since $G \geq (m(J) - \epsilon)N$
(with $\epsilon$ small) it is not hard to show that the Ising system is
above the critical temperature.  Thus, with
large probability, $N_1$ is close to half of $N - G$ -- say
$|2N_1 - (N - G)| < \epsilon N$.  Hence, when all is said and done, we are
reduced
to the problem of the cluster distribution for (subcritical) percolation on the
complete graph with $N_1 \approx (N/2)(1-m)$ sites and bond probability $\approx
J/N$.  Let us temporarily denote these parameters by $n$ and $\alpha/n$ -- with
$\alpha < 1$.  As is not hard to show, the distribution for the cluster size at
any one of these sites is bounded by a Bernoulli branching process with a
mean of
$\alpha$ and a maximum of $n$ offspring.  Thus $F_0 \leq_d I$ where $I$
satisfies the
distributional equality
\begin {equation}
I = 1 + \frac \alpha n\sum_{j=1}^n I^{(j)}
\label{F}
\end{equation}
with the $I^{(j)}$ independent and identical in
distribution to $I$.

Let $g_N(\epsilon)$ denote the probability that
$|G - mN| > \epsilon N$.  Let $\phi_N(\epsilon)$ denote the probability that
$|N_1 - 1/2(N-G)| > \epsilon N$ optimised over all $G$'s such that
$|G - mN| < \epsilon N$.  Then, solving Eq. (\ref{F}) (after expectation)
we find
\begin{equation}
(1 - g_N)(1 - \phi_N)\langle F_0 \rangle_{N,J} \leq
Ng_N + (1-g_N)
\frac {N\phi_N + (1-\phi_N)}{1 - \frac 12 J(1-m) - 2J\epsilon}
\label{upperA}
\end{equation}
Using the large deviation estimate of ~\cite{BGJ} for $g_N$ and a similar
(easily derived) estimate for $\phi_N$ we get, letting $N\to\infty$ and then
$\epsilon\to 0$
\begin{mathletters}
\label{Upp}
\begin {equation}
\langle F_0 \rangle_{J} \leq
\frac {1}{1 - \frac 12 J(1-m)}
\label{up1}
\end{equation}
hence
\begin {equation}
\langle C_0 \rangle_{J} \leq
\frac {1-m}{1 - \frac 12 J(1-m)}.
\label{up2}
\end{equation}
\end{mathletters}

The derivation of the lower bound involves a few more details.  We still
condition on $G \approx mN$ and $N_1 \approx N_2$; now we must pay lip
service to
the possibility of another large cluster.  Let $n$ and $\alpha$ be as before.
We will imagine that there are already $n-1$ sites present and that we add the
$n^{\text{th}}$ at the origin.  Let $c$ satisfy $c^2n = G$.  Then, in order to
get a cluster of size $G$, either (i) the pre-existing collection of $n-1$ sites
must contain a cluster of size larger than $c\sqrt n$ or (ii) the new site must
give rise to at least $c\sqrt n$ bonds.  For ease of future exposition, we will
replace (i) by the weaker condition that some cluster contains at least
$c\sqrt n$
bonds (rather than sites).  Denoting by ${\bf X}_i$ and ${\bf X}_{ii}$ the
indicators of these events, it is not hard to see, by comparison with the
aforementioned branching process, that for the described values of $\alpha$
and $n$,
both probabilities tend to zero at least as fast as $\exp\{-b(\alpha)\sqrt
n\}$ for
some $b > 0$.

Now consider the clusters of the first $n-1$ sites which we denote by
$K_1, K_2, \dots K_s$.  Let $\pi_j$ denote the probability that the new site
connects to the $j^{\text{th}}$ cluster.  Clearly
$p|K_j| \geq \pi_j \geq p|K_j| - p^2|K_j|^2$.  Hence
\begin {equation}
F_0 \geq (1-{\bf X}_i)(1-{\bf X}_{ii})[1 + \sum_{j = 1}^{s}(p|K_j|^2 -
p^2|K_j|^3)]
\end{equation}
The first term in the sum is obviously $p\sum_j F_j$ -- essentially our upper
bound.  But given that ${\bf X}_i \neq 1$, each $|K_j| \leq c\sqrt n$ and thus
\begin {equation}
F_0(N,p_N) \geq
(1-{\bf X}_i)(1-{\bf X}_{ii})
[1 + (1 - \frac {\alpha}{\sqrt n})\frac {\alpha}{n}\sum_j F_j(N-1,p_N).
\label{WC}
\end{equation}
The only remaining difficulty is that the $F's$ on the right hand side of
Eq. (\ref{WC}) are slightly out of balance with regards to their arguments.
However, the density $p_N$ may be obtained from the density $p_{N-1}$ by
independently removing occupied bonds with probability $1/N$.  Thus writing
$F_j(N-1,p_{N}) = F_j(N-1,p_{N-1}) + F_j(N-1,p_{N}) - F_j(N-1,p_{N-1})$, the
difference term is (distributionally) negative and may be bounded below by
$-F_j(N-1,p_{N-1})$ if even a {\em single} bond in the cluster of $j$ gets
removed.  However since we may operate under the stipulation that there are
never
more than $c\sqrt n$ bonds in any of these clusters, the probability of such a
loss is of the order $N^{-1/2}$.  Putting all these ingredients together, we
arrive at the recursive inequality
\begin {equation}
\langle F_0 \rangle_{J,N} \geq
[1 + \frac 12 J(1 - m(J))\langle F_0 \rangle_{J,N}]e(N,\epsilon)
\end{equation}
with $e(N,\epsilon) < 1$ satisfying
$\lim_{\epsilon\to 0}\lim_{N\to\infty}e(N,\epsilon) = 0$  After a
straightforward
limiting argument, the desired result
\begin {equation}
\langle C_0 \rangle_{J} = \frac{1-m}{1 - \frac 12 J(1 - m(J))}
\label{DR}
\end{equation}
now follows from the upper and lower bounds. This gives us $\tilde
\gamma^\prime = 1/2$
for the complete graph.

\subsection {Conclusions/Speculations}
As indicated by our notation the exponents $\tilde \gamma^\prime$, $\tilde
\beta$ and $\tilde \nu^\prime$ have direct counterparts in
spin--systems.  A standard hyper-scaling relation,
$d\gamma^{\prime}/(2 \beta + \gamma^{\prime}) = 2 -\eta$, would therefore
predict
the upper critical dimension $d_c = 6$ (see also Ref. \cite{CL})
This, once again, is
surprising since it differs from the usual value ($d_c = 4$)
associated with Ising systems.

What has so far been demonstrated (in a tautological sense) is a
breakdown in some of the anticipated relations between thermodynamic
and geometric exponents.  Let us illustrate this further.
For magnetic systems, the critical state may be perturbed by a magnetic field
$h$ which serves to define the exponent $\delta$.  This exponent is
related to the geometric exponent $\tau$ by the following
argument:  If
$h\ll1$, only clusters of size on the order of or exceeding $1/h$
will be aligned with the field.  Hence, $m(h) \sim \sum_{n \gtrsim
1/h}P_{n} \sim h^{\tau -2} \equiv h^{1/\delta}$, i.e. $\tau - 2 =
1/\delta$.  This relationship has broken down on the Bethe lattice and
thus it may be presumed not to hold in sufficiently high dimension.  It
therefore must be reinterpreted as a hyperscaling relation.  But there
is a further point to be considered, namely that the above argument must
also break down.

Although this argument is far from rigorous, it
appears to be irrefutable {\em provided that one assumes that the
distribution of clusters with
$h\gtrsim 0$ is not significantly disturbed from the $h = 0$
distribution and that the contribution from the infinite cluster
scales as the contribution from finite clusters}.  We expect that
the breakdown of
$\tau - 2 = 1/\delta$
comes from the fact that the contribution
from finite clusters which scale with an exponent
$\tau -2 = 1/2$ is different from the contribution of the
infinite cluster which
scales with the Ising  exponent $1/\delta= 1/3$.

Let us now present some speculations/conjectures concerning the
behaviour of these systems in finite dimensions.  Under the assumption that
$d_c = 6$,
the exponents $\tilde\nu^{\prime}$, $\tilde\gamma^{\prime}$ etc. calculated here
would be valid for $d \geq 6$ (perhaps with logarithmic modifications at $d
= 6$).
This would
give $\tau_{d = 4} = 7/3$ while $\tau_{d = 6} = 5/2$.  But what about 5
dimensions?  The
simplest scenario is as follows:  Noting that $\delta$ is not a geometric
exponent, let us
eliminate this object in favour of $\eta$ (which is both geometric and
thermodynamic and coincides with the Ising exponent)
via the usual hyperscaling relation.  This gives us the hyperscaling relation
\begin {equation}
\tau = \frac{3d + 2 - \eta}{d + 2 - \eta}
\end{equation}
For $d \geq 4 $ we may set $\eta = 0$. We note that this relation gives the
right value for $d=4$ and $d=6$ and breaks down for $d > 6$ which is consistent
with the upper critical dimensionality $d_c = 6$. We therefore
may suppose the relation to hold for  $d \leq 6$. In particular for
 $d = 5$, it gives us the prediction $\tau_{d = 5} = 17/7$.  Similar
reasoning leads to the predictions for $4 \leq d \leq 6$ of $\tilde
\gamma^\prime = 2/(d-2)$ and $\tilde \nu^\prime = 1/(d-2)$, i.e.
$\tilde \gamma^\prime_{d = 5} = 2/3$ and $\tilde \nu^\prime_{d = 5} =
1/3$.
Both hyperscaling and scaling relations predict $\tilde\alpha^{\prime}
= \frac 12$. On the other hand any {\em thermodynamic} definition
concerning specific heat leads us to $\alpha^{\prime} = 0$. This
dichotomy may be due to the fact that the singularity
$\tilde\alpha^{\prime} = \frac12$ has a vanishing  amplitude.
\cite{CL}

Perhaps the most significant feature uncovered by these calculations
is the appearance of two divergent length scales, $\xi^\prime$
and $\tilde \xi^\prime$ corresponding to the infinite and the finite
clusters.  Let us pause to reinterpret the former.  Note that the
(truncated) correlation function for the probability that two sites
belong to the infinite cluster is also the correlation function for the
probability that the pair does {\it not} belong to the infinite
cluster.  Thus, $\xi^\prime$ can be related to the typical scale of
cavities in the infinite cluster.  In this light, the equality of
$\xi^\prime$ and $\tilde \xi^\prime$ -- as is the case for ordinary
percolation -- is eminently reasonable:  Cavities in the infinite
cluster of sizes up to some scale ($\xi^\prime$) that are populated
with finite cluster that range up to a comparable scale.  However our
situation is quite different.  For $0 < T_c - T \ll 1$ (and $ d > 4$)
we expect the cavities to be filled with relatively small scale finite
clusters.

Finally, let us emphasise the conclusions of the final calculations
in Section B:  The mean field exponents obtained all appear to be
stable to the presence of disorder.  Taken in conjunction with the previous
discussion, -- which includes the assumption/prediction that $d_c = 6$ --
this would
imply a violation of the Harris
criterion--type bound ``$\nu \geq 2/d$'' in dimensions 5--8.  Since the
above bound holds in all systems where it is possible to define an equivalent
{\it finite--size scaling} correlation length,~\cite{CCFS}
the implication here is that it is not possible to define such a length--scale.

 We caution the reader that the discussions in this section are highly
speculative;
the various scenarios might all be wrong.  We are not definitive in any of these
``predictions'' -- they have all been made without the benefit of derivations or
supplementary calculations (let alone rigorous proofs).  However in $d = 5\
\&\ 6$ the
system under consideration is certainly within reach of currently available
numerical
methods which would shed some light on these issues.

\medskip
\noindent One of us (JM) was supported by NSF 96-32898.

\end{document}